\documentclass[aps,floatfix,showpacs,twocolumns,preprintnumbers,amsmath,amssymb]{revtex4}
\usepackage{natbib}
\usepackage{dcolumn}
\usepackage{graphicx}

\newcommand{\be}{\begin{equation}}
\newcommand{\ee}{\end{equation}}
\newcommand{\bea}{\begin{eqnarray}}
\newcommand{\eea}{\end{eqnarray}}
\def\[{\begin{equation}}
\def\]{\end{equation}}
\begin{document}
\title{Luminosity distance and redshift in the Szekeres inhomogeneous cosmological models}
\author{Anthony Nwankwo}
\author{Mustapha Ishak\footnote{Electronic address: mishak@utdallas.edu}}
\author{John Thompson}
\affiliation{
Department of Physics, The University of Texas at Dallas, Richardson, TX 75083, USA}
\date{\today}
\begin{abstract}
The Szekeres inhomogeneous models can be used to model the true lumpy universe that we observe. This family of exact solutions to Einstein's equations was originally derived with a general metric that has no symmetries. In this work, we develop and use a framework to integrate the angular diameter and luminosity distances in the general Szekeres models. We use the affine null geodesic equations in order to derive a set of first-order ordinary differential equations that can be integrated numerically to calculate the partial derivatives of the null vector components. These equations allow the integration in all generality of the distances in the Szekeres models and some examples are given. The redshift is determined from simultaneous integration of the null geodesic equations. This work does not assume spherical or axial symmetry, and the results will be useful for comparisons of the general Szekeres inhomogeneous models to current and future cosmological data.
\end{abstract} 
\maketitle
%
%
\section{Introduction}
%
Recently, there has been a renewed interest in studying some current questions in cosmology using inhomogeneous models, see for example the reviews \cite{Celerier2007,BolejkoEtAlBook} and references therein. Studies have included for example the formation of structures in the universe \cite{KandH,HandK,Bolejko2006a,Bolejko2007} and the question of cosmic acceleration, e.g. see \cite{
Iguchietal2002,Alnesetal2006,Enqvist1,Garfinkle2006,Kaietal2006,Biswasetal2006,TanimotoNambu2007,Hunt,Enqvist,IshakEtAl2008,Chung,Garcia} and references there. It is also of interest that such inhomogeneous models can offer a wider range of interpretation of cosmological observations. 

Whereas one can find a large body of literature studying exact solutions to Einstein's field equations that are inhomogeneous cosmological models \cite{krasinski1997,KSHM}, relatively little work has been done comparing inhomogeneous models to various observations.

In this paper, we consider the calculation of the area and luminosity distances in the general Szekeres models. The affinely parameterized null geodesic equations are used to derive a set of first-order ordinary differential equations ready to be integrated numerically to calculate the partial derivatives of the null vector components. The results allow the integration of the distances in the general Szekeres models. The redshift is   calculated from numerical integration of the set of null geodesics. The general results should be of immediate application to comparisons between the general Szekeres models and cosmological data.

The models were originally derived by Szekeres \cite{Szekeres1,Szekeres2} as an exact solution to Einstein's equations with a general metric that has no symmetries (i.e. no Killing vector fields) and a dust source. The models have been investigated analytically by several authors \cite{Szafron1977,SussmanAndTriginer1999,BonnorAndTomimura1976,Bonnoretal1977,GoodeAndWainwright1982,Bonnor1985,Barrow,SussmanAndTriginer1999,HellabyAndKrasinski2002,Bolejko2006a,Bolejko2007,HellabyAndKrasinski2008,Krasinski2008,Nolan2007}. 
Reference \cite{EllisVanEllst1998} classifies them in the same category as the observed lumpy universe (see the table on page 37 there) and the models are considered as one of the best exact solution candidates to represent the true lumpy universe we live in. The models have been discussed in regards to structure formations in \cite{GoodeAndWainwright1982,Bolejko2006a,Bolejko2007} and also most recently in the context of cosmic acceleration and comparisons to supernova data in \cite{IshakEtAl2008,BolejkoCelerier} and the Rees-Sciama effect in the cosmic microwave background radiation in \cite{Bolejko2009}. 

It is perhaps worth clarifying that we are not proposing the Szekeres model as the true and ultimate model of the universe but rather developing a framework based on exact inhomogeneous cosmological models where cosmological observations can be investigated with a wider range 
of possible interpretations. 
\section{The Szekeres Models}\label{sec:szekeres}
The models have been written in at least three different sets of coordinates: The original set was given by Szekeres in \cite{Szekeres1,Szekeres2}, then Goode and Wainwright \cite{GoodeAndWainwright1982} proposed a second set of coordinates, and then finally a third set of coordinates was used by \cite{Bolejko2006a,Bolejko2007,HellabyAndKrasinski2002}, which we adopt in this paper. The models have been discussed in great detail in \cite{krasinski1997,PlebanskiAndKrasinski2006,HellabyAndKrasinski2008,Krasinski2008} and we give here only a brief introduction to set the notation. 

The Szekeres metric can be written as   
\be
ds^2= -dt^2+\frac{(R_{,r}-R\frac{E_{,r}}{E})^2}{\epsilon-k(r)}dr^2 +\frac{R^2}{E^2}(dp^2+dq^2)
\label{eq:metric}
\ee
where $R=R(t,r)$ is the areal "radius". The function $k(r)$ is related to the energy per unit mass and determines the curvature of the spatial sections $t=constant$. This function divides the models into sub-cases: hyperbolic $(k(r)<0)$, parabolic $(k(r)=0)$, and elliptic $(k(r)>0)$. The function $E=E(r,p,q)$ is given by
\be
E(r,p,q)=\frac{(p^2+q^2)}{2S(r)}-\frac{P(r)}{S(r)}p-\frac{Q(r)}{S(r)}q+C(r)
\ee
and the functions $P(r),S(r),Q(r)$, and $C(r)$ satisfy the relation 
\be
C(r)=\frac{P^2(r)}{2S(r)}+\frac{Q^2(r)}{2S(r)}+\frac{S(r)}{2}\,\epsilon, 
\ee
but are otherwise arbitrary. 

The geometrical constant $\epsilon$ determines whether the $(p,q)$ 2-surfaces are 
spherical ($\epsilon=+1$), pseudo-spherical ($\epsilon=-1$) or planar ($\epsilon=0$).
That is, the constant $\epsilon$ determines how the 2-surfaces of constant $r$ foliate
the 3-dimensional spatial sections of constant $t$. The function $E(r,p,q)$ determines how the coordinates $(p, q)$ are mapped onto the unit 2-sphere, pseudo-sphere or plane for each value of $r$, see for example \cite{HellabyAndKrasinski2002,HellabyAndKrasinski2008,PlebanskiAndKrasinski2006}.

The Einstein field equations with a dust source and no cosmological constant read
\be
(R_{,t}(t,r))^2=\frac{2M(r)}{R(t,r)}-k(r)
\label{eq:EE1}
\ee
and 
\be
8 \pi \rho(t,r,p,q)=\frac{2(M_{,r}(r)-3 M(r) \frac{E_{,r}}{E})}{R^2(R_{,r}-R \frac{E_{,r}}{E})}
\label{eq:EE2}
\ee
where we have set $G=c=1$ and the function $M(r)$ represents the total active gravitational mass in the case $\epsilon =+1$ \cite{HellabyAndKrasinski2002,HellabyAndKrasinski2008}. The evolution of $R(t,r)$ depends on $k(r)$ and is given as follows:

\noindent
The hyperbolic case: $k(r)<0$
\be
R(t,r)=\frac{M(r)}{-k(r)}(\cosh{\eta}-1)
\ee
\be
t-t_B(r)=\sigma\frac{M(r)}{(-k(r))^{3/2}}(\sinh{\eta}-\eta)
\ee
The parabolic case: $k(r)=0$
\be
R(t,r)=M(r) \frac{\eta^2}{2}
\ee
\be
t-t_B(r)=\sigma M(r)\frac{\eta^3}{6}
\ee
The elliptic case: $k(r)>0$
\be
R(t,r)=\frac{M(r)}{k(r)}(1-\cos{\eta})
\ee
\be
t-t_B(r)=\sigma\frac{M(r)}{(k(r))^{3/2}}(\eta-\sin{\eta})
\ee
where $t_B(r)$ is an arbitrary function of $r$ and represents the Big Bang or Crunch time. Also, $\sigma=\pm 1$ is in order to allow for time reversal. There are several sub-cases of the Szekeres models, and a given model is specified by six functions that can be reduced to five by using the coordinate freedom in $r$. This indicates their rich geometry \cite{HellabyAndKrasinski2002,HellabyAndKrasinski2008,Krasinski2008}.
%
\section{A convenient form for the null geodesic equations}
%
The null geodesic equations govern the propagation of light rays in a given spacetime and are necessary to solve in order to derive observable functions for a given model. The non-affinely parameterized null geodesic equations are given by 
\be
k^{\alpha};_{\beta}k^{\beta}=-\frac{1}{\lambda(\tau)} \frac{d\lambda(\tau)}{d\tau}k^{\alpha}
\label{eq:NAgeodesic}
\ee
where $\tau$ is an arbitrary (non-affine) parameter, $\lambda(\tau)$ is a function of $\tau$, and $k^{\alpha}=\frac{dx^{\alpha}}{d\tau}$ is a null tangent vector to the geodesics. 
Now, let's recall that if one applies the following change of parameter (see for example  \cite{PlebanskiAndKrasinski2006}) for a discussion)
\be
\tau \rightarrow s(\tau) = \int^{\tau}_{\tau_0}\frac{C}{\lambda(\underline{\tau})}d\underline{\tau}
\ee
where C is a constant and it follows in this parameterization that 
\be
 {k}^{\alpha};_{\beta} {k}^{\beta}=0 
\label{eq:Ageodesic}
\ee
where the null tangent vector $ {k}^{\alpha}=\frac{dx^{\alpha}}{ds\,\,}$ and the geodesic equations (\ref{eq:Ageodesic}) are all affinely parameterized.

The affinely parameterized null geodesic equations for the Szekeres model are given by \cite{Nolan2007,BolejkoEtAlBook}, (see footnote \cite{NullGeodesicsNote}):

\begin{eqnarray}
&&\dot k^t+ \frac{R,_{tr}-R,_{t}\frac{E,_{r}}{E}}{\epsilon-k}\left(R,_{r}-R\frac{E_{r}}{E}\right)(k^r)^2+\frac{R R,_{t}}{E^2}[(k^p)^2+(k^q)^2]=0 \label{geodkt1}\\ 
&&\dot k^r+2\frac{R,_{tr}-R,_{t}\frac{E,_{r}}{E}}{R,_{r}-R\frac{E,_{r}}{E}}k^t k^r+\left(\frac{R,_{rr}-
R,_{r}\frac{E,_{r}}{E}-R\frac{E,_{rr}}{E}+R(\frac{E,_{r}}{E})^2}{R,_{r}-R\frac{E,_{r}}{E}}+ 
\frac{k,_{r}}{2(\epsilon-k)}\right) (k^r)^2 \nonumber \\
&& +2\frac{R}{E^2}\frac{E,_{r}E,_{p}-E E,_{pr}}{R,_{r}-R\frac{E,_{r}}{E}}k^r k^p+2\frac{R}{E^2}\frac{E,_{r}E,_{q}-E E,_{qr}}{R,_{r}-R\frac{E,_{r}}{E}}k^r k^q
-\frac{R}{E^2}\frac{\epsilon-k}{R,_{r}-R\frac{E,_{r}}{E}}[(k^p)^2+(k^q)^2]=0
\label{geodkr1}\\
&&\dot k^p+2\frac{R,_{t}}{R}k^t k^p -\frac{1}{R}\frac{R,_{r}-R\frac{E,_{r}}{E}}{\epsilon-k}(E,_{r}E,_{p}-EE,_{pr})(k^r)^2 +2\left(\frac{R,_{r}}{R}-\frac{E,_{r}}{E}\right)k^r k^p \nonumber \\
&&-\frac{E,_{p}}{E}(k^p)^2-2\frac{E,_{q}}{E}k^p k^q + \frac{E,_{p}}{E}(k^q)^2=0\label{geodkp1}\\
&&\dot k^q+2\frac{R,_{t}}{R}k^t k^q -\frac{1}{R}\frac{R,_{r}-R\frac{E,_{r}}{E}}{\epsilon-k}(E,_{r}E,_{q}-EE,_{qr})(k^r)^2  +2\left(\frac{R,_{r}}{R}-\frac{E,_{r}}{E}\right)k^r k^q \nonumber \\
&& +\frac{E,_{q}}{E}(k^p)^2-2\frac{E,_{p}}{E}k^p k^q - \frac{E,_{q}}{E}(k^q)^2=0\label{geodkq1}
\end{eqnarray}
where $\dot {}=\frac{d\,\, }{ds}$ and we will use the two notations interchangeably. Now, in order to rewrite these equations in a convenient form for us to solve for the null vector components, we proceed in this way. First, we define the compact functions 
\be
H=\frac{(R_{,r}-R\frac{E_{,r}}{E})^2}{\epsilon-k}
\ee
and 
\be
F=\frac{R}{E}\,\,. 
\ee
The geodesic equation (\ref{geodkt1}) then becomes 
\be
\dot k^t + \frac{1}{2}(H)_{,t}(k^r)^2+\frac{1}{2}(F^2)_{,t}[(k^p)^2+(k^q)^2]=0 \label{geodkt2}.
\ee
We multiply equation (\ref{geodkr1}) by $H$ and use the total derivative expression
\begin{equation}
\dot \Psi=\Psi_{,t}k^t+\Psi_{,r}k^r+\Psi_{,p}k^p+\Psi_{,q}k^q
\label{full2partial}
\end{equation}
in order to rewrite the second geodesic equation (\ref{geodkr1}) as
\be
H\dot k^r+\dot{H}k^r-\frac{1}{2}(H)_{,r}(k^r)^2-\frac{1}{2}(F^2)_{,r}\big{[}(k^p)^2+(k^q)^2\big{]}=0\label{geodkr2}.
\ee
Similarly, we multiply Eqs. (\ref{geodkp1}) and (\ref{geodkq1}) by $F^2$ and use (\ref{full2partial}) to rewrite the third and fourth geodesic equations as 
\be
F^2\dot k^p -\frac{1}{2}(H)_{,p}(k^r)^2+(F^2) {\bf \dot{}} \,\,k^p-\frac{1}{2}(F^2)_{,p}\big{[}(k^p)^2+(k^q)^2\big{]}=0\label{geodkp2}
\ee
\be
F^2\dot k^q -\frac{1}{2}(H)_{,q}(k^r)^2+(F^2){\bf \dot{}} \,\,k^q-\frac{1}{2}(F^2)_{,q}\big{[}(k^p)^2+(k^q)^2\big{]}=0\label{geodkq2}.
\ee
A relation that will be useful for simplifications is the null vector condition $ {k}^{\alpha}  {k}_{\alpha}=0$ and reads 
\begin{equation}
( {k}^t)^2-H( {k}^r)^2-F^2\big{[}( {k}^p)^2+( {k}^q)^2\big{]}=0.
\label{nullaff}
\end{equation}

Now, the null geodesic equations (\ref{geodkt2}), (\ref{geodkr2}), (\ref{geodkp2}), and (\ref{geodkq2}) constitute a system of 4 second-order ordinary differential equations (ODEs) for the functions $\{t(s), r(s), p(s), q(s)\}$ where the coefficients are composed of the metric functions evaluated on the null cone using the field equations and the model specifications. We use a fourth-order Runge-Kutta algorithm with adaptive step size for the numerical integration \cite{Press}. The code iterates between calls to evaluate the field equations on the null cone and calls to integrate the ODEs. We implement the Runge-Kutta code with the function vectors (see for this notation \cite{Press}) given by $\textbf{y}=\{t,r,p,q,\frac{dt}{ds},\frac{dr}{ds},\frac{dp}{ds},\frac{dq}{ds}\}$ and $\frac{d\textbf{y}}{ds}=\{\frac{dt}{ds},\frac{dr}{ds},\frac{dp}{ds},\frac{dq}{ds},\,\frac{d^2t}{ds^2},\frac{d^2r}{ds^2},\frac{d^2p}{ds^2},\frac{d^2q}{ds^2}\}$ so our system of 4 second-order ODEs is transformed into a system of 8 first-order ODEs.

While this integration provides us with the four components $\{k^t,k^r,k^p,k^q\}$  and is enough to compute the redshift, we need to solve for the partial derivatives of these components in order to calculate the area and luminosity distances and we do that in the next two sections. 
%
\section{A set of equations for partial derivatives of the null vector components}
First, we use the null vector condition (\ref{nullaff}) into the r-component of the null geodesic equation (\ref{geodkr2}) to write 
\be
H\frac{d}{ds}(  k^r)+  k^r\frac{d}{ds}H-\frac{1}{2}(H)_{,r}(  k^r)^2-\frac{1}{2}\frac{(F^2)_{,r}}{F^2}\big{[}(  k^t)^2-H(  k^r)^2\big{]}=0\label{geodkr2affine}.
\ee
Next, we note that we can use (\ref{full2partial}) to write the useful relations
\bea
\frac{\partial}{\partial x^{\alpha}}\left[\frac{d(  k^r)}{ds}\right]&=&\frac{\partial}{\partial x^{\alpha}}\left(  k^r_{,\beta}k^{\beta}\right)\nonumber\\
&=&  k^r_{,\beta \alpha}  k^{\beta}+  k^r_{,\beta}  k^{\beta}_{,\alpha}=  k^r_{, \alpha \beta}  k^{\beta}+  k^r_{,\beta}  k^{\beta}_{,\alpha}\nonumber\\
&=&\frac{d}{ds}(  k^r_{,\alpha})+  k^r_{,\beta}  k^{\beta}_{,\alpha},\label{partialdkrds}
\eea
\bea
\frac{\partial}{\partial x^{\alpha}}\left[\frac{d}{ds}H\right]=\frac{\partial}{\partial x^{\alpha}}\left(H_{,\beta}k^{\beta}\right)=H_{,\beta \alpha}  k^{\beta}+H_{,\beta}  k^{\beta}_{,\alpha},\label{partialdHds}
\eea
and 
\be
\frac{\partial}{\partial x^{\alpha}}\left[\frac{d}{ds}(F^2)\right]=\frac{\partial}{\partial x^{\alpha}}\left((F^2)_{,\beta}k^{\beta}\right)=(F^2)_{,\beta \alpha} k^{\beta}+(F^2)_{,\beta} k^{\beta}_{,\alpha}.\label{partialdF2ds}
\ee
Now, taking the partial derivative of equation (\ref{geodkr2affine}) wrt $x^{\alpha}$ and using equations (\ref{partialdkrds}) and (\ref{partialdHds}), we get
\bea
&&H\left[\frac{d}{ds}(  k^r_{,\alpha})+  k^r_{,\beta}  k^{\beta}_{,\alpha}\right]+H_{,\alpha}\frac{d}{ds}(  k^r)
+\left[H_{,\beta \alpha}  k^{\beta}+H_{,\beta}   k^{\beta}_{,\alpha}\right]   k^r + H_{,\beta}  k^{\beta}  k^r_{,\alpha}-\frac{1}{2}H_{,r \alpha}(  k^r)^2-H_{,r}  k^r   k^r_{,\alpha}\nonumber\\
&&-\frac{1}{2}\left(\frac{(F^2)_{,r}}{F^2}\right)_{,\alpha}\left((  k^t)^2-H (  k^r)^2\right)-\frac{(F^2)_{,r}}{F^2}\left(  k^t   k^t_{,\alpha}-\frac{1}{2}H_{,\alpha}(  k^r)^2-H  k^r  k^r_{,\alpha}\right)=0\label{odesystemkr1}
\eea
where $\frac{d}{ds}(k^r)$ is given from equation (\ref{geodkr2affine})
\be
\frac{d}{ds}(  k^r)=-\frac{1}{H}\left[  k^r\frac{d}{ds}H-\frac{1}{2}(H)_{,r}(  k^r)^2-\frac{1}{2}\frac{(F^2)_{,r}}{F^2}\big{[}(  k^t)^2-H(  k^r)^2\big{]}\right].\label{geodkr2affine2}
\ee
In a similar way, we take the partial derivative of the $t$-component of the null geodesic equations, (\ref{geodkt2}), and use the relation (\ref{nullaff}) to  obtain  
\bea
&&\frac{d}{ds}(  k^t_{,\alpha})+  k^t_{,\beta}  k^{\beta}_{,\alpha}+\frac{1}{2}H_{,t \alpha}(  k^r)^2+H_{,t}  k^r   k^r_{,\alpha}+\frac{1}{2}\left(\frac{(F^2)_{,t}}{F^2}\right)_{,\alpha}\left((  k^t)^2-H (  k^r)^2\right)\nonumber\\
&&+\frac{(F^2)_{,t}}{F^2}\left(  k^t   k^t_{,\alpha}-\frac{1}{2}H_{,\alpha}(  k^r)^2-H  k^r  k^r_{,\alpha}\right)=0\label{odesystemkt1}.
\eea
Similarly, from the partial derivatives of the $p$-component and $q$-component of null geodesic equations, i.e. (\ref{geodkp2}) and (\ref{geodkq2}), and using the relation (\ref{partialdF2ds}), we find 
\bea
&&F^2\left[\frac{d}{ds}(  k^p_{,\alpha})+  k^p_{,\beta}  k^{\beta}_{,\alpha}\right]+(F^2)_{,\alpha}\frac{d}{ds}(  k^p)-\frac{1}{2}H_{,p\alpha}(  k^r)^2-H_{,p}  k^r  k^r_{,\alpha}+\left[(F^2)_{,\beta \alpha}  k^{\beta}+(F^2)_{,\beta}  k^{\beta}_{,\alpha}\right]  k^p+\nonumber\\
&&\left((F^2)_{,\beta}  k^{\beta}\right)  k^p_{,\alpha}-\frac{1}{2}\left(\frac{(F^2)_{,p}}{F^2}\right)_{,\alpha}\left((  k^t)^2-H(  k^r)^2\right)-\frac{(F^2)_{,p}}{F^2}\left(  k^t   k^t_{,\alpha}-\frac{1}{2}H_{,\alpha}(  k^r)^2-H  k^r  k^r_{,\alpha}\right)=0\label{odesystemkp1}
\eea
and
\bea
&&F^2\left[\frac{d}{ds}(  k^q_{,\alpha})+  k^q_{,\beta}  k^{\beta}_{,\alpha}\right]+(F^2)_{,\alpha}\frac{d}{ds}(  k^q)-\frac{1}{2}H_{,q\alpha}(  k^r)^2-H_{,q}  k^r  k^r_{,\alpha}+\left[(F^2)_{,\beta \alpha}  k^{\beta}+(F^2)_{,\beta}  k^{\beta}_{,\alpha}\right]  k^q+\nonumber\\
&&\left((F^2)_{,\beta}  k^{\beta}\right)  k^q_{,\alpha}-\frac{1}{2}\left(\frac{(F^2)_{,q}}{F^2}\right)_{,\alpha}\left((  k^t)^2-H(  k^r)^2\right)-\frac{(F^2)_{,q}}{F^2}\left(  k^t   k^t_{,\alpha}-\frac{1}{2}H_{,\alpha}(  k^r)^2-H  k^r  k^r_{,\alpha}\right)=0\label{odesystemkq1}
\eea

Finally, we apply the same process to the null vector condition (\ref{nullaff}) to write.

\be
  k^r_{,\alpha}=\frac{1}{2  k^r}\left[-\frac{H_{,\alpha}}{H}(  k^r)^2+\frac{1}{H}\left(2  k^t  k^t_{,\alpha}-(F^2)_{,\alpha}\left((  k^p)^2+(  k^q)^2\right)-2 F^2\left(  k^p   k^p_{,\alpha}+  k^q   k^q_{,\alpha}\right)\right)\right].
\label{partialforkr}
\ee

Now, equations (\ref{odesystemkr1}), (\ref{odesystemkt1}), (\ref{odesystemkp1}), and (\ref{odesystemkq1}) provide 
a set of 16 first-order ordinary differential equations (4 equations for each $\alpha=t,r,p,q$) that can be integrated numerically for the 16 components $\{k^t_{,\alpha},k^r_{,\alpha},k^p_{,\alpha},k^q_{,\alpha}\}$. This integration is done simultaneously with that of the system from the previous section for the 8 first-order ODEs for the components $k^t$, $k^r$, $k^p$, and $k^q$. The system of the 16 ODEs here is given in the appendix as equations (A1)-(A16).

From a numerical point of view, we found it more practical to solve the system of the 12 ODEs given by equations (\ref{odesystemkt1}), (\ref{odesystemkp1}), (\ref{odesystemkq1}) plus the 4 equations given by (\ref{partialforkr}) (that is the set of equations (A5)-(A20) given in the appendix). Again, we integrate the system of ODEs using a fourth-order Runge-Kutta algorithm with adaptive step size \cite{Press}. The code iterates between calls to evaluate the field equations on the null cone and calls to integrate the ODEs.

With the integration of the null vector components as well as their partial derivatives, we can now calculate the area distance, the redshift, and the luminosity distance. 
%
%
\section{The area and luminosity distances for the Szekeres models}
%
%
As usual, the area distance, $D_A$, is related to the surface area, $\delta S$, of a propagating light front of a bundle of light rays by the relation (see for example \cite{PlebanskiAndKrasinski2006})
\be
\delta S= D^{2}_{A}\delta \Omega
\ee
where $\delta \Omega$ is a solid angle element. 

Now using the relation of the area distance to the expansion optical scalar $\theta$ (see early work by \cite{Sachs,Kantowski} and also \cite{PlebanskiAndKrasinski2006} for a recent discussion), one obtains   
\be
d \ln(\delta S)=2 \theta ds  
\label{eq:D_AToTheta}
\ee
where $s$ is an affine parameter and $\theta$ is given in terms of the affinely parameterized tangent vector as 
\be
\theta=\frac{1}{2}  {k}^\alpha\,_{; \alpha}.
\label{eq:theta}
\ee
It follows from the three above equations that
\be
d \ln{D_A}=\theta\,\,ds=\frac{1}{2} {k}^\alpha\;_{;\alpha}\,\,ds.
\label{eq:AreaDistance}
\ee

The right-hand side integrand of equation (\ref{eq:AreaDistance}) can be evaluated as
\bea
  k^{\alpha}\,_{;\alpha}&=&  k^t_{,t}+  k^r_{,r}+  k^p_{,p}+  k^q_{,q}+  k^t(\Gamma^t_{tt}+\Gamma^r_{tr}+\Gamma^p_{tp}+\Gamma^q_{tq})+  k^r(\Gamma^t_{rt}+\Gamma^r_{rr}+\Gamma^p_{rp}+\Gamma^q_{rq})+\nonumber\\
& &  k^p(\Gamma^t_{pt}+\Gamma^r_{pr}+\Gamma^p_{pp}+\Gamma^q_{pq})+  k^q(\Gamma^t_{qt}+\Gamma^r_{qr}+\Gamma^p_{qp}+\Gamma^q_{qq})\label{D0037}
\eea
where the connection coefficients can be obtained straightforwardly from the metric as:
\begin{eqnarray}
\Gamma^t_{tt}&=&0\nonumber\\ \Gamma^r_{tr}&=&\frac{1}{2}\frac{H_{,t}}{H}\nonumber\\ \Gamma^p_{tp}&=&\frac{1}{2}\frac{(F^2)_{,t}}{F^2}=\Gamma^q_{tq}\nonumber\\ \Gamma^t_{rt}&=&0=\Gamma^t_{pt}=\Gamma^t_{qt}\nonumber\\ \Gamma^r_{rr}&=&\frac{1}{2}\frac{H_{,r}}{H}\nonumber\\ \Gamma^p_{rp}&=&\frac{1}{2}\frac{(F^2)_{,r}}{F^2}=\Gamma^q_{rq}\nonumber\\ \Gamma^r_{pr}&=&\frac{1}{2}\frac{H_{,p}}{H}\nonumber\\  \Gamma^r_{qr}&=&\frac{1}{2}\frac{H_{,q}}{H}\nonumber\\  \Gamma^p_{pp}&=&\frac{1}{2}\frac{(F^2)_{,p}}{F^2}\nonumber\\ \Gamma^q_{qq}&=&\frac{1}{2}\frac{(F^2)_{,q}}{F^2}\nonumber\\  \Gamma^q_{pq}&=&\frac{1}{2}\frac{(F^2)_{,p}}{F^2}\nonumber\\  \Gamma^p_{qp}&=&\frac{1}{2}\frac{(F^2)_{,q}}{F^2}\nonumber \end{eqnarray}

\begin{figure}
\begin{center}
\begin{tabular}{|c|c|}
\hline
{\includegraphics[width=3.in,height=3.in,angle=-90]{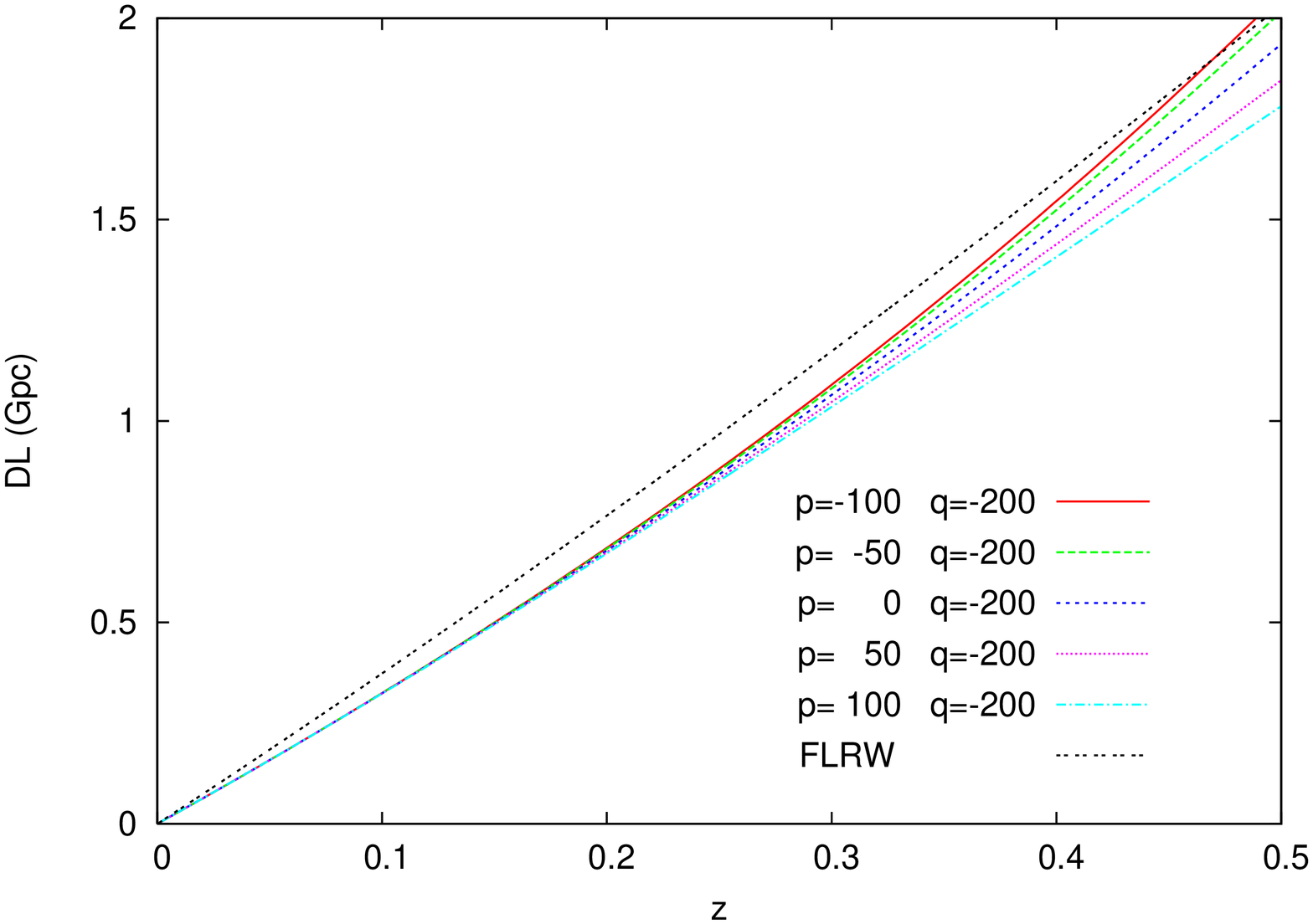}}&
{\includegraphics[width=3.in,height=3.in,angle=-90]{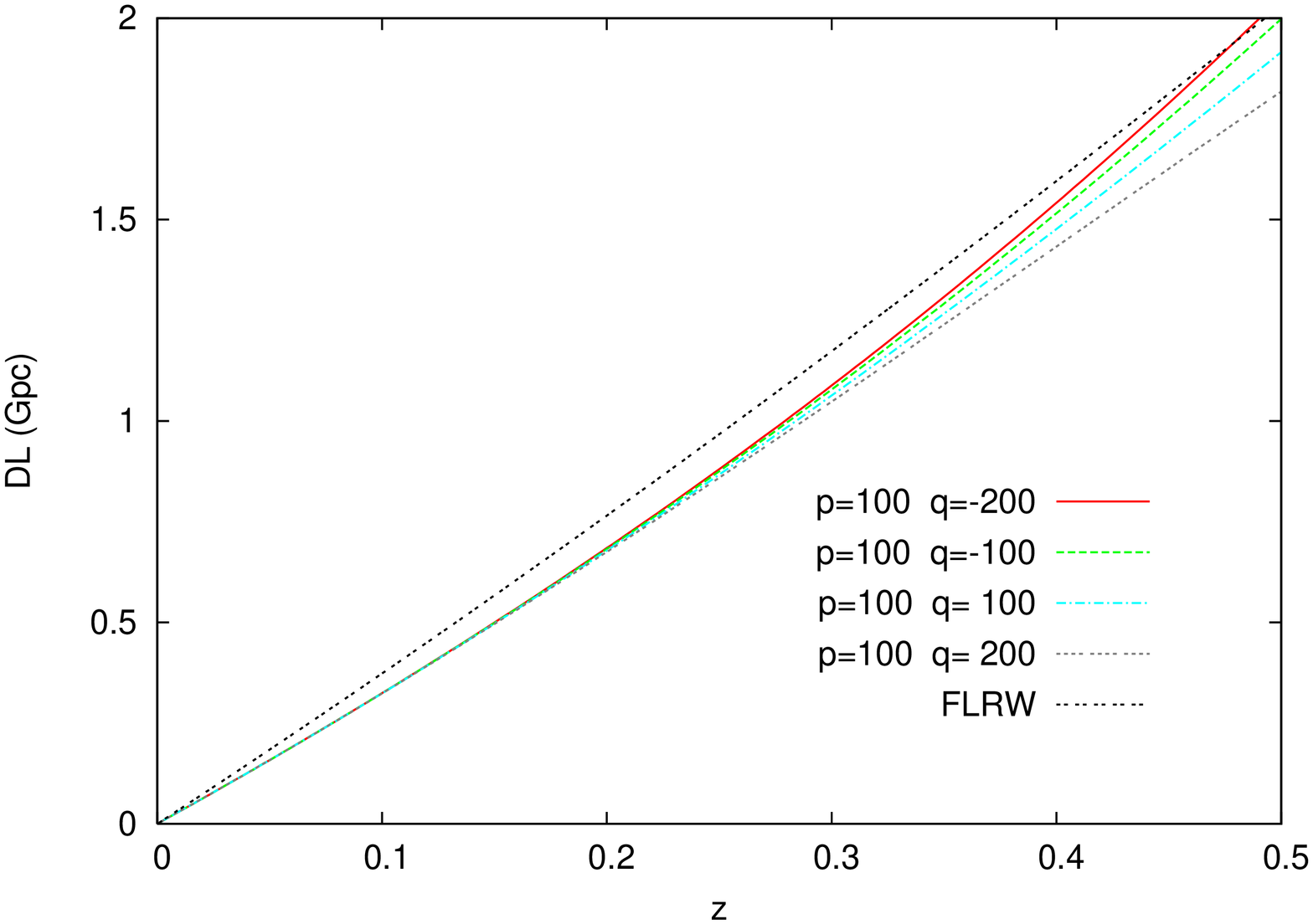}} \\
\hline  
\end{tabular}
\caption{\label{fig:luminosity}
Luminosity distances for a Szekeres model that is not axially or spherically symmetric. To the left, the value of $q$ is fixed to $-200$ while $p$ is varied by taking the values $-100,-50,0,50,100$. To the right, the value of $p$ is fixed to $-100$ while $q$ is varied by taking the values $-200,-100,0,100,200$. The Szekeres inhomogeneous model used here is for an illustration purpose only and was introduced in \cite{Bolejko2006a} within structure formation context. The model is specified in our section V-A. The luminosity distance for an open FLRW model is plotted as well.  
} 
\end{center}
\end{figure}

Putting these into equation (\ref{eq:AreaDistance}) gives 
\bea
d\ln{D_{A}}&=&\Big{[}\frac{1}{2} \left(\frac{(F^2)_{,t}}{F^2}  k^t+\frac{(F^2)_{,r}}{F^2}  k^r+\frac{(F^2)_{,p}}{F^2}  k^p+\frac{(F^2)_{,q}}{F^2}  k^q\right)+\nonumber\\
&&\frac{1}{4}\left(\frac{H_{,t}}{H}  k^t+\frac{H_{,r}}{H}  k^r+\frac{H_{,p}}{H}  k^p+\frac{H_{,q}}{H}  k^q\right)+\frac{1}{2}\left(   k^t_{,t}+  k^r_{,r}+  k^p_{,p}+  k^q_{,q}\right) \Big{]}ds
\label{eq:diffareadistance}
\eea

and upon applying the total derivative equation (\ref{full2partial}), we get the following equations for the area distance:

\begin{equation}
D_A=F\,H^{1/4} \exp{\Big{[}\frac{1}{2} \int^s_{s_0}{\left(  k^t_{,t}+  k^r_{,r}+  k^p_{,p}+  k^q_{,q}\right)} d\underline s\Big{]}}.
\label{AreaDistance2}
\end{equation}
 The luminosity distance is given from its relation to the area distance expression (\ref{AreaDistance2}) as
\be
D_L=(1+z)^2 \,D_A.
\ee
Given explicit functions for a Szekeres model, the area distance is calculated from equation (\ref{AreaDistance2}) where the components
$k^t_{,t}$, $k^r_{,r}$, $k^p_{,p}$, and $k^q_{,q}$ are numerically integrated from the system of ODEs derived in section IV and listed in the Appendix. The redshift is also numerically integrated from the null geodesic equations as described in section VI further.

\subsection{Examples of luminosity distance versus redshift plots}
For illustration, we integrate and plot luminosity distances versus redshift for a 
Szekeres model that is not axially or spherically symmetric. We use a model that was introduced in \cite{Bolejko2006a} within large scale structure formation context and we simply use it here for illustration purposes. In this case, we set:
\begin{itemize}
\item{$\epsilon=+1$.} 
\item{$t_B(r)=0$ for a simultaneous Big Bang.}
\item{$M(r)=(\sinh(r))^3$.}
\item{${k}(r)=\frac{-1}{1+ C^2 r^2}$ (i.e. at large $r$, the spatial curvature goes to zero as indicated by CMB observations \cite{observations5}) and at very small $r$ curvature goes negative in accord with local observations of matter abundances with lower density than the critical density.}
\item{The functions $\{S,P,Q\}$ are given by model-1 of \cite{Bolejko2006a} with $\{140,10, -113\,ln(1+r)\}$.}
\end{itemize}
The results for various values of $p$ and $q$ values are given in Figure I along with an open FLRW model. The exploration of more sophisticated Szekeres models and comparisons to supernova data and other cosmological distances data will be presented in follow-up investigations.  
%
\subsection{Application and verification of special cases}
We derived a general expression for the area distance and luminosity distance for the Szekeres models. It is important to verify that these expressions reduce to ones that we know in the special cases. 
\subsubsection{Axially symmetric case}
As explored in \cite{Nolan2007,BolejkoEtAlBook}, the conditions on the Szekeres metric functions for axial symmetry are given by  
\bea
E E_{,pr}&=&E_{,p}E_{r}\\
E E_{,qr}&=&E_{,q}E_{r}.
\label{AxialSymcondition}
\eea

It is trivial to see from our compact notation in the geodesic equations (\ref{geodkp2}) and (\ref{geodkq2}) that these axial symmetry conditions are here simply 
\be
H_{,p}=H_{,q}=0.
\label{AxialSymCondOurNotation}
\ee
Conditions (\ref{AxialSymCondOurNotation}) assure that $dq=dp=0$ holds along the whole geodesic. We proceed then by setting $dp=dq=0$ into equation (\ref{nullaff}) so $(k^t)^2-H(  k^r)^2=0$ or 
\be
(  k^r)^2=\frac{1}{H}(  k^t)^2.
\label{nullaxial}
\ee
Using this in equation (\ref{geodkt2}) and integrating for $(  k^t)^2$ gives
\be
(  k^t)^2=\exp{\left(-\int^t_{t_0}{\frac{H_{,\underline t}}{H}}d\underline t\right)}\label{ktaxial}
\ee
Now, from (\ref{nullaxial}), we have  
\be
\frac{dt}{dr}=-\sqrt{H} 
\ee
and using it in equation (\ref{ktaxial}) and taking its square root gives
\be
 {k^t}=\exp{\left(\int^r_{r_0}{(\sqrt{H}),_t}d\underline r \right)}
\label{as:kt}
\ee
where we have also noted that $\frac{1}{2}\frac{H_{,t}}{\sqrt{H}}=(\sqrt{H})_{,t}$.

Next, using (\ref{nullaxial}), we get immediately for $ {k^r}$:
\bea
 {k^r}=-\frac{1}{\sqrt{H}}\exp{\left(\int^r_{r_0}{(\sqrt{H}),_t}d\underline r \right)}
\label{as:kr}
\eea
These two expressions agree with the ones obtained in \cite{BolejkoEtAlBook} for the axially symmetric case.

Now, our expression (\ref{eq:diffareadistance}) for the area distance reduces to
\be
d\ln{D_{A}}=\Big{[}\frac{1}{2} \left(\frac{(F^2)_{,t}}{F^2}  k^t+\frac{(F^2)_{,r}}{F^2}  k^r\right)+
\frac{1}{4}\left(\frac{H_{,t}}{H}  k^t+\frac{H_{,r}}{H}  k^r\right)+\frac{1}{2}\left(   k^t_{,t}+  k^r_{,r}\right) \Big{]}ds
\label{eq:diffareadistanceAxial}
\ee
where we use the tangent vector component equations (\ref{as:kt}) and (\ref{as:kr}) for axial symmetry to express the partial derivatives in the  last two terms in (\ref{eq:diffareadistanceAxial}) as 
\bea
  k^t_{,t}&=&\Big{[}-\frac{1}{2}\int^t_{t_0}{\frac{\left(H_{,\underline t}\right)}{H}}d\underline t\Big{]}_{,t}  k^t\label{axialkt}\\
  k^r_{,r}&=&-\frac{1}{2}\frac{H_{,r}}{H}  k^r+\Big{[}-\frac{1}{2}\int^t_{t_0}{\frac{\left(H_{,\underline t}\right)}{H}}d\underline t\Big{]}_{,r}  k^r\label{axialkr}.
\eea
and where we have used $\frac{dr}{dt}dt=-\frac{1}{\sqrt{H}}dt$ and $\frac{1}{2}\frac{H_{,t}}{\sqrt{H}}=(\sqrt{H})_{,t}$ in equation(\ref{axialkr}). We also note that with axial symmetry $p$ and $q$ are being constant and we can combine the term in (\ref{axialkt}) plus the last term in (\ref{axialkr}) to make a full derivative. Putting this into the integration of the last part of equation (\ref{eq:diffareadistanceAxial}) gives 
\bea
\frac{1}{2}\int^s_{s_0} \left(  k^t_{,t}+  k^r_{,r}\right) d\underline s &=& 
-\frac{1}{4}
\int^r_{r_0} \frac{H_{,\underline r}}{H} d\underline r- \frac{1}{4} \int^t_{t_0} \frac{H_{,\underline t}}{H} d\underline t
\label{axialDa3}.
\eea
These two terms cancel with the integrated two middle terms in equation (\ref{eq:diffareadistanceAxial}) yielding a straightforward integration result 
\be
D_A=F.
\label{eq:DAaxial}
\ee
This result is in agreement with the result of \cite{BolejkoPrivate,BolejkoCelerier} of the axially symmetric case. 
%
\subsubsection{Spherically symmetric case}
%
%
In the more special case of spherical symmetry, i.e. $\epsilon=+1$ and $E_{,r}=0$,  our expression (\ref{eq:diffareadistance}) for the area distance reduces to simply 
\be
d\ln{D_A}=\frac{1}{2}\left(\frac{(R^2)_{,t}}{R^2}  k^t+\frac{(R^2)_{,r}}{R^2}  k^r\right)ds.
\ee
Integrating both sides gives
\be
D_A=R
\ee
which is the well-known result for the observer area distance for an observer located at the center of the Lemaitre-Tolman-Bondi spherical model, see for example \cite{PlebanskiAndKrasinski2006}.
%
\section{The redshift in the Szekeres models}
%
In order to plot the luminosity distance as a function of the redshift, the latter needs also to be integrated in all generality. We start with the standard relation 
\begin{equation}
1+z=\frac{(  k^\alpha u_{\alpha})_e}{(  k^\alpha u_{\alpha})_o}=\frac{(  k^t)_e}{(  k^t)_o}
\label{RedshiftStandard}
\end{equation}
where $  k^{\alpha}$ is the affinely parameterized null vector, $u^{\alpha}=(1,0,0,0)$ is the 4-velocity vector, and the subscripts $e$ and $o$ are for emitted (at the source) and observed (at the observer) respectively.  Now, we use equation (\ref{geodkt2}) and write 
\bea
\frac{d[(  k^t)^2]}{ds}=-(H)_{,t}  k^t(  k^r)^2-\frac{(F^2)_{,t}}{F^2}  k^t[(  k^t)^2-H(  k^r)^2]=2  k^t \frac{d  k^t}{ds}\label{diffkt}
\eea
which we use along with taking the natural log of both sides of (\ref{RedshiftStandard}) and differentiating both sides wrt to the affine parameter $s$ to obtain
\begin{equation}
\frac{1}{1+z}\frac{d z}{d s}=\frac{1}{  k^t}\frac{d  k^t}{ds}=-\frac{1}{2\,  k^t}\left[(H)_{,t}(  k^r)^2+\frac{(F^2)_{,t}}{F^2}((  k^t)^2-H(  k^r)^2)\right].
\label{eq:SzekeresRedshift}
\end{equation}

While it is informative to see this expression, which one can integrate simultaneously
with the three other null geodesic equations (\ref{geodkr2}), (\ref{geodkp2}), and (\ref{geodkq2}) to find
the redshift, one can equally use equation (\ref{RedshiftStandard}) and get $k^t$ by integrating
the four null geodesic equations. We proceeded with the latter method for our numerical calculations. An alternative way to find the redshift and the related drift effects in the Szekeres models can be found in \cite{KrasinskiandBolejko2011}.

As an aside, we note that using the null condition $  k^\alpha   k_{\alpha}=0$ to substitute $( {k}^t)^2-H( {k}^r)^2$ by $F^2[( {k}^p)^2+( {k}^q)^2]$, the redshift equation (\ref{eq:SzekeresRedshift}) can be written as  
\begin{equation}
\frac{1}{1+z}\frac{d z}{d s}=-\frac{1}{2\,  k^t}\left[(H)_{,t}(  k^r)^2+(F^2)_{,t}((  k^p)^2+(  k^q)^2)\right]\label{szekeresredshift}
\end{equation} 
which is precisely the redshift relation that was derived in our previous work \cite{IshakEtAl2008,FootnoteRedshift}.
%
\subsection{Application and verification of special cases}
We derived a general expression for the redshift for the Szekeres models. It is important to verify that this expression reduces to the ones in the known special cases.
\subsubsection{Axially symmetric case}
In their recent book \cite{BolejkoEtAlBook}, the authors derived an expression for the redshift for the axially symmetric Szekeres models that reads: 
\begin{equation}
\ln{(1+z)}=\int^{r_0}_{r_e}{dr\frac{R_{,tr}-R_{,t}\frac{E_{,r}}{E}}{\sqrt{1-k}}}.
\label{bolejko(3.100)}
\end{equation}
Taking the derivative of this equation with respect to the affine parameter, $s$, we get
\begin{eqnarray}
\frac{1}{1+z}\frac{d(1+z)}{ds}&=&\frac{R_{,tr}-R_{,t}\frac{E_{,r}}{E}}{\sqrt{1-k}}\frac{dr}{ds}
\label{modbolejko1}
\end{eqnarray}
Using our equation for the null condition (\ref{nullaff}) and $  k^p=  k^q=0$ gives
\begin{equation}
\frac{  k^r}{  k^t}=-\frac{\sqrt{1-k}}{R_{,r}-R \frac{E_{,r}}{E}}
\label{modbolejko2}
\end{equation}
Now we multiply equation (\ref{modbolejko1}) by $\frac{dr/ds}{dr/ds}=\frac{  k^r}{  k^r}$ to write
\begin{eqnarray}
\frac{1}{1+z}\frac{d(1+z)}{ds}&=&\frac{R_{,tr}-R_{,t}\frac{E_{,r}}{E}}{\sqrt{1-k}}\frac{(  k^r)^2}{  k^r}.
\label{modbolejko11}
\end{eqnarray}
We then use equation(\ref{modbolejko2}) to substitute for $  k^r$ in the denominator of equation (\ref{modbolejko11}) and expand the numerator to get
\begin{equation}
\frac{1}{1+z}\frac{d(1+z)}{ds}=-\frac{1}{  k^t}\frac{R_{,tr}R_{,r}+R R_{,t}(\frac{E_{,r}}{E})^2-(R_{,r}R_{,t}+R R_{,tr})\frac{E_{,r}}{E}}{1-k}(  k^r)^2
\label{bolejko(3.100)b}
\end{equation}
Now, we go back to our expression for the redshift, i.e. equation (\ref{szekeresredshift}) and we  put there $  k^p=  k^q=0$ as well as the definitions of $H$ and $F$ to re-write it as follows
\begin{equation}
\frac{1}{1+z}\frac{d(1+z)}{ds}=-\frac{1}{  k^t}\frac{R_{,tr}R_{,r}+R R_{,t}(\frac{E_{,r}}{E})^2-(R_{,r}R_{,t}+R R_{,tr})\frac{E_{,r}}{E}}{1-k}(  k^r)^2
\label{ourredshift}\end{equation}
As one can see our expression for the redshift when reduced to the axially symmetric case agrees with the redshift given in the work of \cite{BolejkoEtAlBook} for this particular case.
%
\subsubsection{Spherically symmetric case}
The spherically symmetric case happens in the more particular case where $E_{,r}=0$. Substituting this condition into equation (\ref{ourredshift}) gives 
\begin{equation}
\frac{1}{1+z}\frac{d(1+z)}{ds}=-\frac{1}{  k^t}\frac{R_{,tr}R_{,r}}{1-k}(  k^r)^2.
\label{redshiftreduce}\end{equation}
Using equation (\ref{modbolejko2}) into equation (\ref{redshiftreduce})
gives 
\begin{equation}
\frac{1}{1+z}\frac{d(1+z)}{ds}=\frac{R_{,tr}}{\sqrt{1-k}}   k^r
\end{equation}
that upon integration yields
\begin{equation}
\ln{(1+z)}=\int^{r_0}_{r_e}{dr\frac{R_{,tr}}{\sqrt{1-k}}}
\end{equation}
which is the usual result for the spherically symmetric Lemaitre-Tolman-Bondi models, see for example  \cite{PlebanskiAndKrasinski2006}.
%
\section{Conclusion}
We derived and used a framework to integrate the area and luminosity distances in the general Szekeres models. We used the general affinely parameterized null geodesic equations in order to derive a set of first-order ordinary differential equations that can be integrated numerically to calculate the partial derivatives of the null vector components. These equations allow the numerical integration of the area and luminosity distances in the general Szekeres models.  
We determined the redshift from simultaneous integration of the null geodesic equations. This work does not assume spherical or axial symmetry and will be useful for comparisons of the general Szekeres inhomogeneous models to current and future cosmological data.
\acknowledgments 
We thank A. Krasinski, K. Bolejko and M-N. Celerier for useful comments on our previous work on this topic. We thank K. Bolejko for useful comments about the special axially symmetric case. We thank Jason Dossett and Austin Peel for reading the paper. MI acknowledges that this material is based upon work supported in part by NASA under grant NNX09AJ55G. 
%
\appendix
%
%
\section{Set of ODEs for the partial derivatives of the null vector components}
Equation (\ref{odesystemkr1}) from section IV expands into the following  4 ODEs:
%
\bea
&&H\left[\frac{d}{ds}(  k^r_{,t})+  k^r_{,t}  k^{t}_{,t}+  k^r_{,r}  k^{r}_{,t}+  k^r_{,p}  k^{p}_{,t}+  k^r_{,q}  k^{q}_{,t}\right]+H_{,t}\frac{d}{ds}(  k^r)\nonumber\\
&&+\left[H_{,t t}  k^{t}+H_{,r t}  k^{r}+H_{,p t}  k^{p}+H_{,q t}  k^{q}+H_{,t}   k^{t}_{,t}+H_{,r}   k^{r}_{,t}+H_{,p}   k^{p}_{,t}+H_{,q}   k^{q}_{,t}\right]   k^r +\nonumber\\
&& H_{,t}  k^{t}  k^r_{,t}+H_{,r}  k^{r}  k^r_{,t}+H_{,p}  k^{p}  k^r_{,t}+H_{,q}  k^{q}  k^r_{,t}-\frac{1}{2}H_{,r t}(  k^r)^2-H_{,r}  k^r   k^r_{,t}\nonumber\\
&&-\frac{1}{2}\left(\frac{(F^2)_{,r}}{F^2}\right)_{,t}\left((  k^t)^2-H (  k^r)^2\right)-\frac{(F^2)_{,r}}{F^2}\left(  k^t   k^t_{,t}-\frac{1}{2}H_{,t}(  k^r)^2-H  k^r  k^r_{,t}\right)=0\eea
\bea
&&H\left[\frac{d}{ds}(  k^r_{,r})+  k^r_{,t}  k^{t}_{,r}+  k^r_{,r}  k^{r}_{,r}+  k^r_{,p}  k^{p}_{,r}+  k^r_{,q}  k^{q}_{,r}\right]+H_{,r}\frac{d}{ds}(  k^r)\nonumber\\
&&+\left[H_{,t r}  k^{t}+H_{,r r}  k^{r}+H_{,p r}  k^{p}+H_{,q r}  k^{q}+H_{,t}   k^{t}_{,r}+H_{,r}   k^{r}_{,r}+H_{,p}   k^{p}_{,r}+H_{,q}   k^{q}_{,r}\right]   k^r +\nonumber\\
&& H_{,t}  k^{t}  k^r_{,r}+H_{,r}  k^{r}  k^r_{,r}+H_{,p}  k^{p}  k^r_{,r}+H_{,q}  k^{q}  k^r_{,r}-\frac{1}{2}H_{,r r}(  k^r)^2-H_{,r}  k^r   k^r_{,r}\nonumber\\
&&-\frac{1}{2}\left(\frac{(F^2)_{,r}}{F^2}\right)_{,r}\left((  k^t)^2-H (  k^r)^2\right)-\frac{(F^2)_{,r}}{F^2}\left(  k^t   k^t_{,r}-\frac{1}{2}H_{,r}(  k^r)^2-H  k^r  k^r_{,r}\right)=0\eea
\bea
&&H\left[\frac{d}{ds}(  k^r_{,p})+  k^r_{,t}  k^{t}_{,p}+  k^r_{,r}  k^{r}_{,p}+  k^r_{,p}  k^{p}_{,p}+  k^r_{,q}  k^{q}_{,p}\right]+H_{,p}\frac{d}{ds}(  k^r)\nonumber\\
&&+\left[H_{,t p}  k^{t}+H_{,r p}  k^{r}+H_{,p p}  k^{p}+H_{,q p}  k^{q}+H_{,t}   k^{t}_{,p}+H_{,r}   k^{r}_{,p}+H_{,p}   k^{p}_{,p}+H_{,q}   k^{q}_{,p}\right]   k^r +\nonumber\\
&& H_{,t}  k^{t}  k^r_{,p}+H_{,r}  k^{r}  k^r_{,p}+H_{,p}  k^{p}  k^r_{,p}+H_{,q}  k^{q}  k^r_{,p}-\frac{1}{2}H_{,r p}(  k^r)^2-H_{,r}  k^r   k^r_{,p}\nonumber\\
&&-\frac{1}{2}\left(\frac{(F^2)_{,r}}{F^2}\right)_{,p}\left((  k^t)^2-H (  k^r)^2\right)-\frac{(F^2)_{,r}}{F^2}\left(  k^t   k^t_{,p}-\frac{1}{2}H_{,p}(  k^r)^2-H  k^r  k^r_{,p}\right)=0\eea
\bea
&&H\left[\frac{d}{ds}(  k^r_{,q})+  k^r_{,t}  k^{t}_{,q}+  k^r_{,r}  k^{r}_{,q}+  k^r_{,p}  k^{p}_{,q}+  k^r_{,q}  k^{q}_{,q}\right]+H_{,q}\frac{d}{ds}(  k^r)\nonumber\\
&&+\left[H_{,t q}  k^{t}+H_{,r q}  k^{r}+H_{,p q}  k^{p}+H_{,q q}  k^{q}+H_{,t}   k^{t}_{,q}+H_{,r}   k^{r}_{,q}+H_{,p}   k^{p}_{,q}+H_{,q}   k^{q}_{,q}\right]   k^r +\nonumber\\
&& H_{,t}  k^{t}  k^r_{,q}+H_{,r}  k^{r}  k^r_{,q}+H_{,p}  k^{p}  k^r_{,q}+H_{,q}  k^{q}  k^r_{,q}-\frac{1}{2}H_{,r q}(  k^r)^2-H_{,r}  k^r   k^r_{,q}\nonumber\\
&&-\frac{1}{2}\left(\frac{(F^2)_{,r}}{F^2}\right)_{,q}\left((  k^t)^2-H (  k^r)^2\right)-\frac{(F^2)_{,r}}{F^2}\left(  k^t   k^t_{,q}-\frac{1}{2}H_{,q}(  k^r)^2-H  k^r  k^r_{,q}\right)=0
\eea
where $\frac{d}{ds}(k^r)$ is written in terms of metric functions and null vector components from equation (\ref{geodkr2affine}) (i.e. (\ref{geodkr2affine2})). 

Equation (\ref{odesystemkt1}) from section IV expands into the following  4 ODEs:
\bea
&&\frac{d}{ds}(  k^t_{,t})+  k^t_{,t}  k^{t}_{,t}+  k^t_{,r}  k^{r}_{,t}+  k^t_{,p}  k^{p}_{,t}+  k^t_{,q}  k^{q}_{,t}+ \frac{1}{2}H_{,t t}(  k^r)^2+H_{,t}  k^r   k^r_{,t}+\frac{1}{2}\left(\frac{(F^2)_{,t}}{F^2}\right)_{,t}\left((  k^t)^2-H (  k^r)^2\right)\nonumber\\
&&+\frac{(F^2)_{,t}}{F^2}\left(  k^t   k^t_{,t}-\frac{1}{2}H_{,t}(  k^r)^2-H  k^r  k^r_{,t}\right)=0
\eea
\bea
&&\frac{d}{ds}(  k^t_{,r})+  k^t_{,t}  k^{t}_{,r}+  k^t_{,r}  k^{r}_{,r}+  k^t_{,p}  k^{p}_{,r}+  k^t_{,q}  k^{q}_{,r}+ \frac{1}{2}H_{,t r}(  k^r)^2+H_{,t}  k^r   k^r_{,r}+\frac{1}{2}\left(\frac{(F^2)_{,t}}{F^2}\right)_{,r}\left((  k^t)^2-H (  k^r)^2\right)\nonumber\\
&&+\frac{(F^2)_{,t}}{F^2}\left(  k^t   k^t_{,r}-\frac{1}{2}H_{,r}(  k^r)^2-H  k^r  k^r_{,r}\right)=0
\eea
\bea
&&\frac{d}{ds}(  k^t_{,p})+  k^t_{,t}  k^{t}_{,p}+  k^t_{,r}  k^{r}_{,p}+  k^t_{,p}  k^{p}_{,p}+  k^t_{,q}  k^{q}_{,p}+ \frac{1}{2}H_{,t p}(  k^r)^2+H_{,t}  k^r   k^r_{,p}+\frac{1}{2}\left(\frac{(F^2)_{,t}}{F^2}\right)_{,p}\left((  k^t)^2-H (  k^r)^2\right)\nonumber\\
&&+\frac{(F^2)_{,t}}{F^2}\left(  k^t   k^t_{,p}-\frac{1}{2}H_{,p}(  k^r)^2-H  k^r  k^r_{,p}\right)=0
\eea
\bea
&&\frac{d}{ds}(  k^t_{,q})+  k^t_{,t}  k^{t}_{,q}+  k^t_{,r}  k^{r}_{,q}+  k^t_{,p}  k^{p}_{,q}+  k^t_{,q}  k^{q}_{,q}+ \frac{1}{2}H_{,t q}(  k^r)^2+H_{,t}  k^r   k^r_{,q}+\frac{1}{2}\left(\frac{(F^2)_{,t}}{F^2}\right)_{,q}\left((  k^t)^2-H (  k^r)^2\right)\nonumber\\
&&+\frac{(F^2)_{,t}}{F^2}\left(  k^t   k^t_{,q}-\frac{1}{2}H_{,q}(  k^r)^2-H  k^r  k^r_{,q}\right)=0.
\eea

Equation (\ref{odesystemkp1}) from section IV expands into the following  4 ODEs:
\bea
&&F^2\left[\frac{d}{ds}(  k^p_{,t})+  k^p_{,t}  k^{t}_{,t}+  k^p_{,r}  k^{r}_{,t}+  k^p_{,p}  k^{p}_{,t}+  k^p_{,q}  k^{q}_{,t}\right]+(F^2)_{,t}\frac{d}{ds}(  k^p)-
\frac{1}{2}H_{,pt}(  k^r)^2-H_{,p}  k^r  k^r_{,t}+\nonumber\\
&&\left[(F^2)_{,t t}  k^{t}+(F^2)_{,r t}  k^{r}+(F^2)_{,p t}  k^{p}+(F^2)_{,q t}  k^{q} +(F^2)_{,t}  k^{t}_{,t}+(F^2)_{,r}  k^{r}_{,t} +(F^2)_{,p}  k^{p}_{,t}+(F^2)_{,q}  k^{q}_{,t}\right]  k^p+\nonumber\\
&&\left((F^2)_{,t}  k^{t}+(F^2)_{,r}  k^{r}+(F^2)_{,p}  k^{p}+(F^2)_{,q}  k^{q}\right)  k^p_{,t}-\frac{1}{2}\left(\frac{(F^2)_{,p}}{F^2}\right)_{,t}\left((  k^t)^2-H(  k^r)^2\right)\nonumber\\
&&-\frac{(F^2)_{,p}}{F^2}\left(  k^t   k^t_{,t}-\frac{1}{2}H_{,t}(  k^r)^2-H  k^r  k^r_{,t}\right)=0
\eea
\bea
&&F^2\left[\frac{d}{ds}(  k^p_{,r})+  k^p_{,t}  k^{t}_{,r}+  k^p_{,r}  k^{r}_{,r}+  k^p_{,p}  k^{p}_{,r}+  k^p_{,q}  k^{q}_{,r}\right]+(F^2)_{,r}\frac{d}{ds}(  k^p)-
\frac{1}{2}H_{,pr}(  k^r)^2-H_{,p}  k^r  k^r_{,r}+\nonumber\\
&&\left[(F^2)_{,t r}  k^{t}+(F^2)_{,r r}  k^{r}+(F^2)_{,p r}  k^{p}+(F^2)_{,q r}  k^{q} +(F^2)_{,t}  k^{t}_{,r}+(F^2)_{,r}  k^{r}_{,r} +(F^2)_{,p}  k^{p}_{,r}+(F^2)_{,q}  k^{q}_{,r}\right]  k^p+\nonumber\\
&&\left((F^2)_{,t}  k^{t}+(F^2)_{,r}  k^{r}+(F^2)_{,p}  k^{p}+(F^2)_{,q}  k^{q}\right)  k^p_{,r}-\frac{1}{2}\left(\frac{(F^2)_{,p}}{F^2}\right)_{,r}\left((  k^t)^2-H(  k^r)^2\right)\nonumber\\
&&-\frac{(F^2)_{,p}}{F^2}\left(  k^t   k^t_{,r}-\frac{1}{2}H_{,r}(  k^r)^2-H  k^r  k^r_{,r}\right)=0
\eea
\bea
&&F^2\left[\frac{d}{ds}(  k^p_{,p})+  k^p_{,t}  k^{t}_{,p}+  k^p_{,r}  k^{r}_{,p}+  k^p_{,p}  k^{p}_{,p}+  k^p_{,q}  k^{q}_{,p}\right]+(F^2)_{,p}\frac{d}{ds}(  k^p)-
\frac{1}{2}H_{,pp}(  k^r)^2-H_{,p}  k^r  k^r_{,p}+\nonumber\\
&&\left[(F^2)_{,t p}  k^{t}+(F^2)_{,r p}  k^{r}+(F^2)_{,p p}  k^{p}+(F^2)_{,q p}  k^{q} +(F^2)_{,t}  k^{t}_{,p}+(F^2)_{,r}  k^{r}_{,p} +(F^2)_{,p}  k^{p}_{,p}+(F^2)_{,q}  k^{q}_{,p}\right]  k^p+\nonumber\\
&&\left((F^2)_{,t}  k^{t}+(F^2)_{,r}  k^{r}+(F^2)_{,p}  k^{p}+(F^2)_{,q}  k^{q}\right)  k^p_{,p}-\frac{1}{2}\left(\frac{(F^2)_{,p}}{F^2}\right)_{,p}\left((  k^t)^2-H(  k^r)^2\right)\nonumber\\
&&-\frac{(F^2)_{,p}}{F^2}\left(  k^t   k^t_{,p}-\frac{1}{2}H_{,p}(  k^r)^2-H  k^r  k^r_{,p}\right)=0
\eea
\bea
&&F^2\left[\frac{d}{ds}(  k^p_{,q})+  k^p_{,t}  k^{t}_{,q}+  k^p_{,r}  k^{r}_{,q}+  k^p_{,p}  k^{p}_{,q}+  k^p_{,q}  k^{q}_{,q}\right]+(F^2)_{,q}\frac{d}{ds}(  k^p)-
\frac{1}{2}H_{,pq}(  k^r)^2-H_{,p}  k^r  k^r_{,q}+\nonumber\\
&&\left[(F^2)_{,t q}  k^{t}+(F^2)_{,r q}  k^{r}+(F^2)_{,p q}  k^{p}+(F^2)_{,q q}  k^{q} +(F^2)_{,t}  k^{t}_{,q}+(F^2)_{,r}  k^{r}_{,q} +(F^2)_{,p}  k^{,p}_{q}+(F^2)_{,q}  k^{q}_{,q}\right]  k^p+\nonumber\\
&&\left((F^2)_{,t}  k^{t}+(F^2)_{,r}  k^{r}+(F^2)_{,p}  k^{p}+(F^2)_{,q}  k^{q}\right)  k^p_{,q}-\frac{1}{2}\left(\frac{(F^2)_{,p}}{F^2}\right)_{,q}\left((  k^t)^2-H(  k^r)^2\right)\nonumber\\
&&-\frac{(F^2)_{,p}}{F^2}\left(  k^t   k^t_{,q}-\frac{1}{2}H_{,q}(  k^r)^2-H  k^r  k^r_{,q}\right)=0
\eea
where $\frac{d}{ds}(k^p)$ is written in terms of the metric functions and the null vector components from equation (\ref{geodkp2}). 

Equation (\ref{odesystemkq1}) from section IV expands into the following  4 ODEs:
\bea
&&F^2\left[\frac{d}{ds}(  k^q_{,t})+  k^q_{,t}  k^{t}_{,t}+  k^q_{,r}  k^{r}_{,t}+  k^q_{,p}  k^{p}_{,t}+  k^q_{,q}  k^{q}_{,t}\right]+(F^2)_{,t}\frac{d}{ds}(  k^q)-
\frac{1}{2}H_{,qt}(  k^r)^2-H_{,q}  k^r  k^r_{,t}+\nonumber\\
&&\left[(F^2)_{,t t}  k^{t}+(F^2)_{,r t}  k^{r}+(F^2)_{,p t}  k^{p}+(F^2)_{,q t}  k^{q} +(F^2)_{,t}  k^{t}_{,t}+(F^2)_{,r}  k^{r}_{,t} +(F^2)_{,p}  k^{p}_{,t}+(F^2)_{,q}  k^{q}_{,t}\right]  k^q+\nonumber\\
&&\left((F^2)_{,t}  k^{t}+(F^2)_{,r}  k^{r}+(F^2)_{,p}  k^{p}+(F^2)_{,q}  k^{q}\right)  k^q_{,t}-\frac{1}{2}\left(\frac{(F^2)_{,q}}{F^2}\right)_{,t}\left((  k^t)^2-H(  k^r)^2\right)\nonumber\\
&&-\frac{(F^2)_{,q}}{F^2}\left(  k^t   k^t_{,t}-\frac{1}{2}H_{,t}(  k^r)^2-H  k^r  k^r_{,t}\right)=0
\eea
\bea
&&F^2\left[\frac{d}{ds}(  k^q_{,r})+  k^q_{,t}  k^{t}_{,r}+  k^q_{,r}  k^{r}_{,r}+  k^q_{,p}  k^{p}_{,r}+  k^q_{,q}  k^{q}_{,r}\right]+(F^2)_{,r}\frac{d}{ds}(  k^q)-
\frac{1}{2}H_{,qr}(  k^r)^2-H_{,q}  k^r  k^r_{,r}+\nonumber\\
&&\left[(F^2)_{,t r}  k^{t}+(F^2)_{,r r}  k^{r}+(F^2)_{,p r}  k^{p}+(F^2)_{,q r}  k^{q} +(F^2)_{,t}  k^{t}_{,r}+(F^2)_{,r}  k^{r}_{,r} +(F^2)_{,p}  k^{p}_{,r}+(F^2)_{,q}  k^{q}_{,r}\right]  k^q+\nonumber\\
&&\left((F^2)_{,t}  k^{t}+(F^2)_{,r}  k^{r}+(F^2)_{,p}  k^{p}+(F^2)_{,q}  k^{q}\right)  k^q_{,r}-\frac{1}{2}\left(\frac{(F^2)_{,q}}{F^2}\right)_{,r}\left((  k^t)^2-H(  k^r)^2\right)\nonumber\\
&&-\frac{(F^2)_{,q}}{F^2}\left(  k^t   k^t_{,r}-\frac{1}{2}H_{,r}(  k^r)^2-H  k^r  k^r_{,r}\right)=0
\eea
\bea
&&F^2\left[\frac{d}{ds}(  k^q_{,p})+  k^q_{,t}  k^{t}_{,p}+  k^q_{,r}  k^{r}_{,p}+  k^q_{,p}  k^{p}_{,p}+  k^q_{,q}  k^{q}_{,p}\right]+(F^2)_{,p}\frac{d}{ds}(  k^q)-
\frac{1}{2}H_{,qp}(  k^r)^2-H_{,q}  k^r  k^r_{,p}+\nonumber\\
&&\left[(F^2)_{,t p}  k^{t}+(F^2)_{,r p}  k^{r}+(F^2)_{,p p}  k^{p}+(F^2)_{,q p}  k^{q} +(F^2)_{,t}  k^{t}_{,p}+(F^2)_{,r}  k^{r}_{,p} +(F^2)_{,p}  k^{p}_{,p}+(F^2)_{,q}  k^{q}_{,p}\right]  k^q+\nonumber\\
&&\left((F^2)_{,t}  k^{t}+(F^2)_{,r}  k^{r}+(F^2)_{,p}  k^{p}+(F^2)_{,q}  k^{q}\right)  k^q_{,p}-\frac{1}{2}\left(\frac{(F^2)_{,q}}{F^2}\right)_{,p}\left((  k^t)^2-H(  k^r)^2\right)\nonumber\\
&&-\frac{(F^2)_{,q}}{F^2}\left(  k^t   k^t_{,p}-\frac{1}{2}H_{,p}(  k^r)^2-H  k^r  k^r_{,p}\right)=0
\eea
\bea
&&F^2\left[\frac{d}{ds}(  k^q_{,q})+  k^q_{,t}  k^{t}_{,q}+  k^q_{,r}  k^{r}_{,q}+  k^q_{,p}  k^{p}_{,q}+  k^q_{,q}  k^{q}_{,q}\right]+(F^2)_{,q}\frac{d}{ds}(  k^q)-
\frac{1}{2}H_{,qq}(  k^r)^2-H_{,q}  k^r  k^r_{,q}+\nonumber\\
&&\left[(F^2)_{,t q}  k^{t}+(F^2)_{,r q}  k^{r}+(F^2)_{,p q}  k^{p}+(F^2)_{,q q}  k^{q} +(F^2)_{,t}  k^{t}_{,q}+(F^2)_{,r}  k^{r}_{,q} +(F^2)_{,p}  k^{p}_{,q}+(F^2)_{,q}  k^{q}_{,q}\right]  k^q+\nonumber\\
&&\left((F^2)_{,t}  k^{t}+(F^2)_{,r}  k^{r}+(F^2)_{,p}  k^{p}+(F^2)_{,q}  k^{q}\right)  k^q_{,q}-\frac{1}{2}\left(\frac{(F^2)_{,q}}{F^2}\right)_{,q}\left((  k^t)^2-H(  k^r)^2\right)\nonumber\\
&&-\frac{(F^2)_{,q}}{F^2}\left(  k^t   k^t_{,q}-\frac{1}{2}H_{,q}(  k^r)^2-H  k^r  k^r_{,q}\right)=0.
\eea
where $\frac{d}{ds}(k^q)$ is written in terms of the metric functions and the null vector components from equation (\ref{geodkq2}). 

Finally, equation (\ref{partialforkr}) from section IV expands into the following  4 equations:
\bea
&&  k^r_{,t}=\frac{1}{2  k^r}\left[-\frac{H_{,t}}{H}(  k^r)^2+\frac{1}{H}\left(2  k^t  k^t_{,t}-(F^2)_{,t}\left((  k^p)^2+(  k^q)^2\right)-2 F^2\left(  k^p   k^p_{,t}+  k^q   k^q_{,t}\right)\right)\right]
\label{partialforkrt}
\eea
\bea
&&  k^r_{,r}=\frac{1}{2  k^r}\left[-\frac{H_{,r}}{H}(  k^r)^2+\frac{1}{H}\left(2  k^t  k^t_{,r}-(F^2)_{,r}\left((  k^p)^2+(  k^q)^2\right)-2 F^2\left(  k^p   k^p_{,r}+  k^q   k^q_{,r}\right)\right)\right]
\label{partialforkrr}
\eea
\bea
&&  k^r_{,p}=\frac{1}{2  k^r}\left[-\frac{H_{,p}}{H}(  k^r)^2+\frac{1}{H}\left(2  k^t  k^t_{,p}-(F^2)_{,p}\left((  k^p)^2+(  k^q)^2\right)-2 F^2\left(  k^p   k^p_{,p}+  k^q   k^q_{,p}\right)\right)\right]
\label{partialforkrp}
\eea
\bea
&&  k^r_{,q}=\frac{1}{2  k^r}\left[-\frac{H_{,q}}{H}(  k^r)^2+\frac{1}{H}\left(2  k^t  k^t_{,q}-(F^2)_{,q}\left((  k^p)^2+(  k^q)^2\right)-2 F^2\left(  k^p   k^p_{,q}+  k^q   k^q_{,q}\right)\right)\right]
\eea
%
%

\end{document}